\documentclass{optica-article}

\journal{opticajournal} 

\articletype{Research Article}

\usepackage{lineno}

\begin{document}

\title{The Surface Orientation Ambiguity for Single Molecules at Dielectric Interfaces}

\author{E. Dey,\authormark{1,$\dag$} M. Elorza,\authormark{2,3,$\dag$,*} F.W. Foss,\authormark{4,$\dag$} J.J. Gomez Cadenas,\authormark{2,$\dag$} and B.J.P. Jones\authormark{1,$\dag$}}

\address{\authormark{1}Department of Physics, University of Texas at Arlington, Arlington, TX 76019, USA\\
\authormark{2}Donostia International Physics Center, BERC Basque Excellence Research Centre, Manuel de Lardizabal 4, San Sebasti\'an / Donostia, E-20018, Spain\\
\authormark{3}Department of Physics, Universidad del Pa\'is Vasco (EHU), Bilbao, E-48080, Spain\\
\authormark{4}Department of Chemistry and Biochemistry, University of Texas at Arlington, Arlington, TX 76019, USA\\

\authormark{$\dag$}The authors contributed equally to this work.\\
\email{\authormark{*}mikel.elorza@dipc.org}} 


\begin{abstract}
    Fluorescent molecules emit light in a dipole radiation pattern that can be used to infer their orientation through defocused fluorescence microscopy.  Proper measurement of the orientation requires mathematical modeling of the radiation pattern expected for a dipole in the geometry of interest, and subsequent comparison against experimental data.  We point out an ambiguity in common calculations of these patterns that appears to compromise orientation measurements for molecules that are especially near to dielectric surfaces.  This results in a rotation of the measured emission dipole toward the surface for near-interface molecules, which can be mistaken for a preferentially horizontal orientation among the emitters.  The proper treatment for on-surface emitters requires consideration of finite-sized current elements between two dielectric media, and we show that the theoretical ambiguity can be lifted via finite-element modeling. A prescription is provided for correcting measured orientations at arbitrary interfaces.  
\end{abstract}

\section{Introduction}

\begin{figure}
\begin{centering}
\includegraphics[width=1\columnwidth]{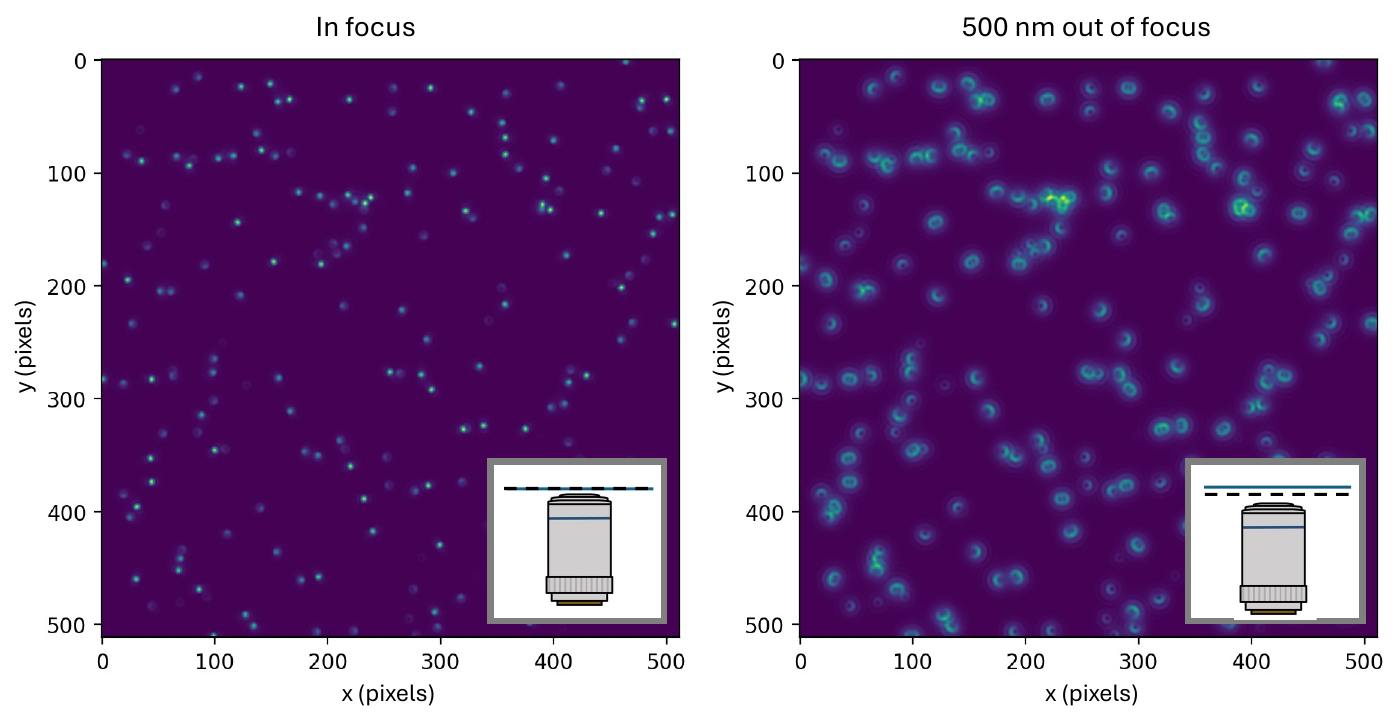}
\par\end{centering}
\caption{Simulated images of in-focus (left) and defocused (right) single molecule fluorescence microscope images. In the defocused image, the orientations of individual molecules become resolvable, in principle, through analysis of their dipole emission patterns. }\label{fig:Examples-of-templates}
\end{figure}

The measurement of molecular orientation using defocused fluorescence
microscopy is a widely used and well documented approach in single
molecule fluorescence imaging. Individual point-like dipole emitters
resolve as individual dots when the molecules lie within the microscope
focal plane (an example simulated image for a typical in-focus microscope
configuration is shown in Fig. \ref{fig:Examples-of-templates}, left). If the emitters lie slightly outside the focal plane, for example,
if the microscope objective is pulled backward from the in-focus position
by around one light wavelength, each emitter projects a more complex
pattern on the image plane (Fig. \ref{fig:Examples-of-templates},
right). The
individual patterns encode information about the emission dipole of each
source molecule. Their analysis can thus be used to infer
the orientation of each molecule in the image, producing potentially
useful structural information about the way molecules are arranged
within a material, inside a cell, or on a substrate. 

Defocused imaging has become a powerful tool in various fields of research, providing critical insights into a wide range of molecular and structural phenomena. This technique is extensively applied to study transition dipoles in molecules and quantum dots~\cite{SCHUSTER2005quantumdots,Debarre2004DipoleOrientation,Bartko1999Orientations}, investigate structural heterogeneities in lipids and complex systems~\cite{Lu2020SMOLM,Dedecker2009Defocused}, and analyze molecular distributions in polymer systems~\cite{Deres2011Heterogeneity,Bartko2002GlassyPolymer}. Additionally, it has been used to characterize growth of materials for organic light-emitting diodes (OLEDs)~\cite{TenopalaCarmona2023Orientation} and elucidate molecular dynamics at surfaces and within polymers~\cite{Hutchison2014surfacebound,Melnikov2007donoracceptor}. Defocused imaging also aids in understanding metal-induced energy transfer~\cite{Ghosh2019grapheneMIET}, donor-acceptor emission in Förster resonance energy transfer (FRET)~\cite{Melnikov2007donoracceptor}, and plasmon-induced effects~\cite{Moon2023Defocused,Zelger2020Defocused}, with ever-expanding applications across both fundamental and applied scientific disciplines.

Various algorithms exist for extracting molecular orientation from
defocused fluorescence data. In principle, the problem can be treated
 analytically by analyzing the radial and azimuthal moments of the molecule image. This approach becomes
more difficult if the patterns from more than one molecule overlap. Another popular method is to simulate pre-calculated
emission patterns associated with various emission dipole angles,
and perform a pixel-wise likelihood maximization. In this way, the
optimal reconstructed polar $\Theta$ and $\Phi$ coordinates for each molecule orientation
can be obtained. The extraction of molecular orientations using this
 method relies on having a suitably accurate prediction of the
image templates to use in the fit. This demands not only a precise
statement of imaging parameters such as defocus distance, system magnification,
emission wavelength, refractive indices of materials, and pupil size,
but also a robust mathematical model of the expected shape of the
collected image given these parameters. 

The calculations used to produce such templates have become canonical
in the field, and are well documented \cite{Novotny_Hecht_2012}. The method is based on calculation of the electric fields in the far-field of oscillating electric
dipole radiating in the vicinity of a dielectric interface. The basic elements of this classical electromagnetism problem date back to Sommerfeld
\cite{Sommerfeld}, and have been further developed by numerous authors since. An authoritative series of works by Lukosz ~\cite{Lukosz_III} and Lukosz
and Kunz ~\cite{Lukosz_kunz_I,Lukosz_kunz_II} present a comprehensive discussion of this physical
system. The textbook by Novotny and Hecht presents the calculation
in detail and also discusses its use in nanophotonics problems \cite{Novotny_Hecht_2012}.
We briefly summarize this ``standard approach'' in the following.

We consider an interface between two media with refractive indices $n_1$ and $n_2$, where $n_1$ lies in the direction of the imaging plane. We must calculate the observed image pattern for an emitter either above or below this surface.  In both cases, the first step is to obtain a ``primary'' electric field $\mathbf{E}_{p}$ from solution
of Maxwells equations, assuming the dipole is
situated in an infinite extent of a single material. This field can
be expressed in the angular spectrum representation, whereby the radiation
field is decomposed in terms of individual propagating and evanescent
waves with specific wave vectors $\mathbf{k}$, which can be either real or imaginary. In order to solve
Maxwells equations for a dipole sitting near a dielectric interface between two media,
these angular components can each be superposed with a wave in
the same half-space as the dipole (the reflected wave), and another
wave in the opposite half-space (the transmitted wave). The amplitude
of these two additional waves is determined by the well-known Fresnel reflection
and electromagnetic transmission coefficients, generating a reflected
electric field $\mathbf{E}_{r}$ and a transmitted field $\mathbf{E}_{t}$,
both expressed in the angular spectrum representation. By the
principle of superposition, it is also admissible to add to this electric
field configuration any homogeneous solution $\mathbf{E}_{0}$ - that
is, any configuration of electric fields that satisfies the Maxwells
equations with no currents or charges in this geometry. As such, the
total electric field is

\begin{equation}
\mathbf{E}(\mathbf{r})=\mathbf{E}_{0}(\mathbf{r})+\mathbf{E}_{p}(\mathbf{r})+\mathbf{E}_{r}(\mathbf{r})+\mathbf{E}_{t}(\mathbf{r}).\label{eq:GreensFunction}
\end{equation}
All of $\mathbf{E}_{p}$, $\mathbf{E}_{r}$ and $\mathbf{E}_{t}$
are sourced by the oscillating electric dipole. Due to the linearity
of Maxwells equations, the dipole-sourced fields can be expressed
in terms of a dyadic Greens function $\overleftrightarrow{\mathbf{G}}(\mathbf{r},\mathbf{r}')$

\begin{equation}
\mathbf{E}_{p}(\mathbf{r})+\mathbf{E}_{r}(\mathbf{r})+\mathbf{E}_{t}(\mathbf{r})=i\omega\mu_{0}\mu\int_{V}\overleftrightarrow{\mathbf{G}}(\mathbf{r},\mathbf{r}')\mathbf{j}(\mathbf{r}')dV',
\end{equation}
where $\overleftrightarrow{\mathbf{G}}(\mathbf{r},\mathbf{r}')=[\mathbf{G}_{x}(\mathbf{r},\mathbf{r}'),\mathbf{G}_{y}(\mathbf{r},\mathbf{r}'),\mathbf{G}_{z}(\mathbf{r},\mathbf{r}')]$
and the vector function $\mathbf{G}_{i}(\mathbf{r},\mathbf{r}')$
represents the vector electric field at position $\mathbf{r}$ generated
by a point current element at $\mathbf{r}'$ that points in the $i$ direction. The integral
is over the volume $V$ where the currents are contained, and $\mathbf{E}(\mathbf{r})$ is predicted everywhere outside of this volume. The above calculation
is exact. It predicts some some non-trivial effects that have been
pointed out by various authors, e.g.~\cite{novotny1997allowed,Novotny_Hecht_2012}. For dipoles within a
few wavelengths of the interface, for example, the near-field components
of the dipole emission pattern that would not ordinarily propagate
as radiation can be scattered into the propagating regime by the presence
of the interface. 

If all that is of interest is the pattern observed in the imaging
system, it is not necessary to become concerned with such near-field
details. The important information is encoded in the electric
field at the back focal plane, which is in the far-field and
can be obtained through the dyadic Greens function evaluated at that
location, $\overleftrightarrow{\mathbf{G}}_{bfp}$. This Greens function
depends on the pixel position in the back focal plane (represented by
2D polar coordinates $\rho,\phi$), dipole unit vector $\mathbf{p}$,
and the distance from the emitter to the interface $\Delta$. For a single dipole with unit
vector $\mathbf{p}$ positioned $r=0$, the electric field at the
back focal plane $\mathbf{E}_{bfp}$ for an in-focus emitter is obtained from the Greens function
$\overleftrightarrow{\mathbf{G}}_{bfp}$ via
\begin{equation}
\mathbf{E}_{bfp}(\phi,\rho,\Delta)=\overleftrightarrow{\mathbf{G}}_{bfp}(\phi,\rho,\Delta)\mathbf{p},
\end{equation}
where $\overleftrightarrow{\mathbf{G}}_{bfp}$ depends ~\cite{Moer}
on three functions $c_{1},c_{2},c_{3}$, as
\begin{equation}
\mathbf{\overleftrightarrow{G}}_{bfp}(\phi,\rho,\Delta)=\left[\begin{array}{ccc}
c_{3}(\rho)\sin^{2}\phi+c_{2}(\rho)\cos^{2}\phi\sqrt{1-\rho^{2}} & -\sin(2\phi)(c_{3}(\rho)-c_{2}(\rho)\sqrt{1-\rho^{2}})/2 & -c_{1}(\rho)\rho\cos\phi\\
-\sin(2\phi)(c_{3}(\rho)-c_{2}(\rho)\sqrt{1-\rho^{2}})/2 &c_{3}(\rho)\cos^{2}\phi+c_{2}(\rho)\sin^{2}\phi\sqrt{1-\rho^{2}} & -c_{1}(\rho)\rho\sin\phi\\
0 & 0 & 0
\end{array}\right]\label{eq:GreensFunction-1}
\end{equation}

\begin{figure}
\begin{centering}
\includegraphics[width=1\columnwidth]{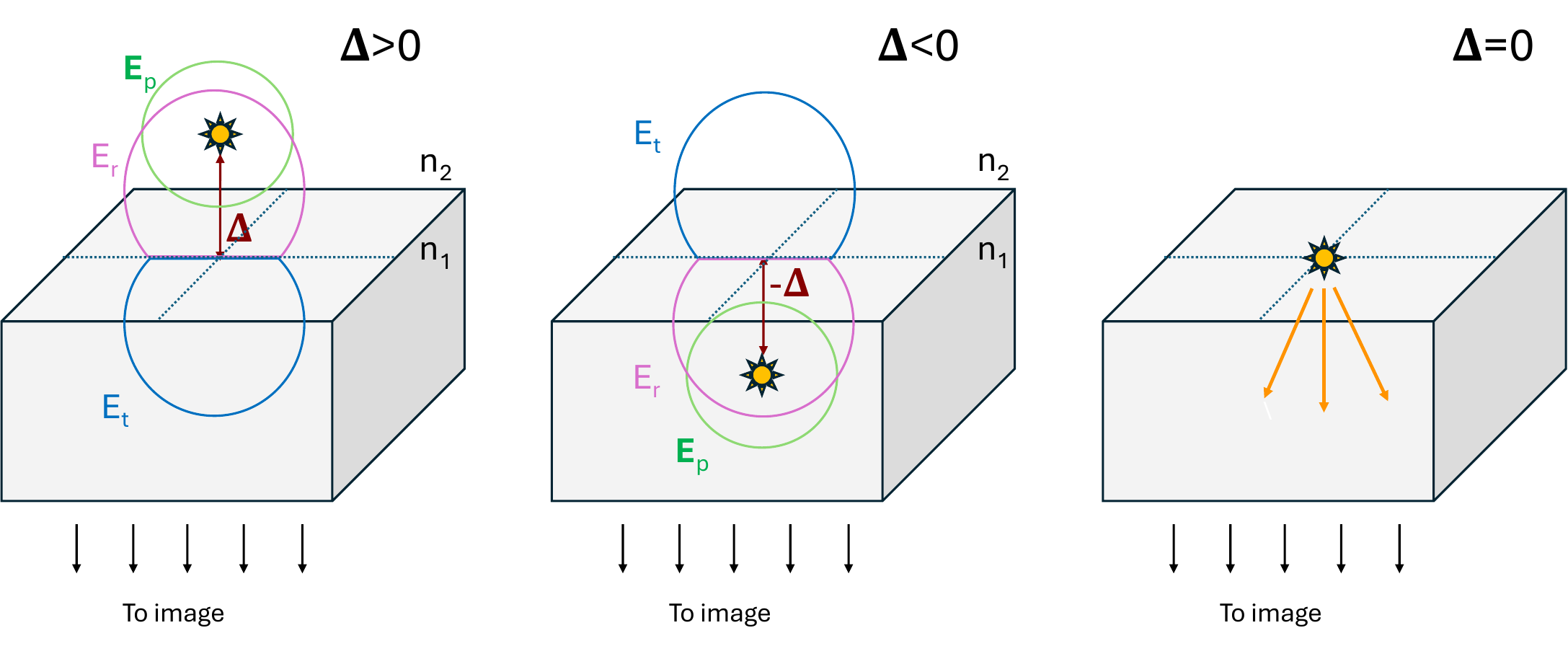}
\par\end{centering}

\caption{Illustration of the geometries considered in this work. In the left figure, the emitter is situated above the interface and the transmitted field is detected; in the central figure, the emitter is below the interface and the primary and reflected  fields are detected.  In the right figure, the molecule is situated at the interface, in contact with both media, and the proper prescription appears to become ambiguous. }\label{fig:Geometries}

\end{figure}

The electric field for an out-of-focus emitter can then be obtained from the above via addition of a phase factor,
\begin{equation}
\mathbf{E}_{bfp}^{defocus}=e^{in_1kd(1-\rho^2)^{1/2}}\mathbf{E}_{bfp},
\end{equation}
where $d$ is the defocus distance. The $c_{i}(\rho)$ functions have distinct forms if the emitter is placed
above or below the interface with respect to the imaging system. The
case $\Delta>0$ corresponds to the ``above'' interface case, where
the transmitted field component is the one observed at the imaging
system. The case $\Delta<0$ corresponds to a below-interface dipole,
where it is the sum of the primary and reflected fields that are transported
to the imaging system in the far field. These example cases are shown
in Fig \ref{fig:Geometries}, left and center. Identical functions
$c_{i}(\rho)$ imply identical values for the intensity distribution
at the back focal plane and hence identical microscope images. As
such, evaluation of the image shape amounts largely to evaluating
the functions $c_{i}(\rho)$ and propagating these through the backfocal
plane Greens function, to generate the microscope image. 

The electric field in the focal plane can be easily obtained by as long as the inclination of the light rays emerging from the tube lens of the microscope is small enough for the paraxial theory of Fourier optics is valid. This is generally true, as the focal length of the tube lens is usually much larger than the focal length of the high NA microscope objectives used for these applications. Thus at the image plane,
\begin{equation}
    \mathbf{E}_{img}(\phi',\rho',\Delta)=C\mathcal{F}\{\mathbf{E}_{bfp}(\phi,\rho,\Delta)\},
\end{equation}
where $\mathcal{F}$ is the Fourier transform operator, $C$ is a phase factor accounting for the waveform correction, and $\phi'$ and $\rho'$ are the coordinates in the image plane~\cite{goodman2005introduction}.

For the placement above the interface $\Delta>0$ (Fig \ref{fig:Geometries},
left) the emitter is located within volume 2 and it is the light transmitted
into volume 1 that is detected, and the $c$ coefficients are given by

\begin{equation}
c_{1}^{\Delta>0}=\left(\frac{n_{1}}{n_{2}}\right)^{2}\frac{\cos\theta_{1}}{\cos\theta_{2}}t_{2\rightarrow1}^{p}(\theta)e^{ik\Delta n_{2}\left[1-(n_{1}/n_{2}\rho)^{2}\right]^{1/2}},
\end{equation}

\begin{equation}
c_{2}^{\Delta>0}=\left(\frac{n_{1}}{n_{2}}\right)t_{2\rightarrow1}^{p}(\theta)e^{ik\Delta n_{2}\left[1-(n_{1}/n_{2}\rho)^{2}\right]^{1/2}},
\end{equation}

\begin{equation}
c_{3}^{\Delta>0}=\left(\frac{n_{1}}{n_{2}}\right)\frac{\cos\theta_{1}}{\cos\theta_{2}}t_{2\rightarrow1}^{s}(\theta)e^{ik\Delta n_{2}\left[1-(n_{1}/n_{2}\rho)^{2}\right]^{1/2}}.
\end{equation}

For placement below the interface, $\Delta<0$ (Fig \ref{fig:Geometries},
center) the molecule is located in volume 1 and it is the primary
and reflected light that is detected and the $c$ coefficients take the
form

\begin{equation}
c_{1}^{\Delta<0}=e^{-ik|\Delta|n_{1}(1-\rho)^{1/2}}+r_{1\rightarrow1}^{p}(\theta)e^{ik|\Delta|n_{1}(1-\rho^{2})^{1/2}},
\end{equation}

\begin{equation}
c_{2}^{\Delta<0}=e^{-ik|\Delta|n_{1}(1-\rho)^{1/2}}-r_{1\rightarrow1}^{p}(\theta)e^{ik|\Delta|n_{1}(1-\rho^{2})^{1/2}},
\end{equation}

\begin{equation}
c_{3}^{\Delta<0}=e^{-ik|\Delta|n_{1}(1-\rho)^{1/2}}+r_{1\rightarrow1}^{s}(\theta)e^{ik|\Delta|n_{1}(1-\rho^{2})^{1/2}}.
\end{equation}

The angles $\theta_{1}$ and $\theta_{2}$ are related by Snell's law
$n_{1}\sin\theta_{1}=n_{2}\sin\theta_{2}$, and $\theta_{1}$ maps
onto a radial coordinate on the back focal plane of the microscope
objective via $\theta_{1}=\sin^{-1}\rho$. In these expressions, $r_{a\rightarrow b}^{s}$,
$r_{a\rightarrow b}^{p}$, $t_{a\rightarrow b}^{s}$, $t_{a\rightarrow b}^{p}$
are the Fresnel coefficients for s-polarized and p-polarized waves,
respectively, which are given in terms of the relevant angles and
refractive indices via

\begin{equation}
r_{1\rightarrow1}^{s}=\frac{n_{1}\cos\theta_{1}-n_{2}\cos\theta_{2}}{n_{1}\cos\theta_{1}+n_{2}\cos\theta_{2}},\quad r_{1\rightarrow1}^{p}=\frac{n_{2}\cos\theta_{1}-n_{1}\cos\theta_{2}}{n_{2}\cos\theta_{1}+n_{1}\cos\theta_{2}}
\end{equation}

\begin{equation}
t_{2\rightarrow1}^{p}=\frac{2n_{2}\cos\theta_{2}}{n_{1}\cos\theta_{2}+n_{2}\cos\theta_{1}},\quad t_{2\rightarrow1}^{s}=\frac{2n_{2}\cos\theta_{2}}{n_{2}\cos\theta_{2}+n_{1}\cos\theta_{1}}
\end{equation}

These exact analytic equations are fairly simple, and can be evaluated with minimal computational load. This has made them a
convenient enabling ingredient in a large number of data published
data analyses. One frequent use case is to model and then measure
the orientation of molecules suspended in a biological sample. This
may commonly correspond to the case of $\Delta<0$, with the molecule
somewhere within a sample that is in contact above with an air interface. Analogues of these calculations for more complex layered media have
also been developed~\cite{lukosz1981light}.

The class of experiments that concern us in this paper are those that
involve the measurement of molecules that are affixed directly to
a surface (Fig \ref{fig:Geometries}, right). Experiments of this
class include Refs~\cite{Hutchison2014surfacebound,TenopalaCarmona2023Orientation}. This corresponds to the limit $\Delta\rightarrow0^{+}$,
and a typical assumption is to consider the relevant short-distance
limit of the above-surface solution to predict the emission patterns
for these molecules.

\begin{equation}
c_{1}^{\Delta\rightarrow0^{+}}=\left(\frac{n_{1}}{n_{2}}\right)^{2}\frac{2n_{2}\cos\theta_{1}}{n_{1}\cos\theta_{2}+n_{2}\cos\theta_{1}},\quad\quad c_{2}^{\Delta\rightarrow0^{+}}=\frac{2n_{1}\cos\theta_{2}}{n_{1}\cos\theta_{2}+n_{2}\cos\theta_{1}},\quad\quad c_{3}^{\Delta\rightarrow0^{+}}=\frac{2n_{1}\cos\theta_{1}}{n_{2}\cos\theta_{2}+n_{1}\cos\theta_{1}}.
\end{equation}

\begin{figure}
\begin{centering}
\includegraphics[width=1\columnwidth]{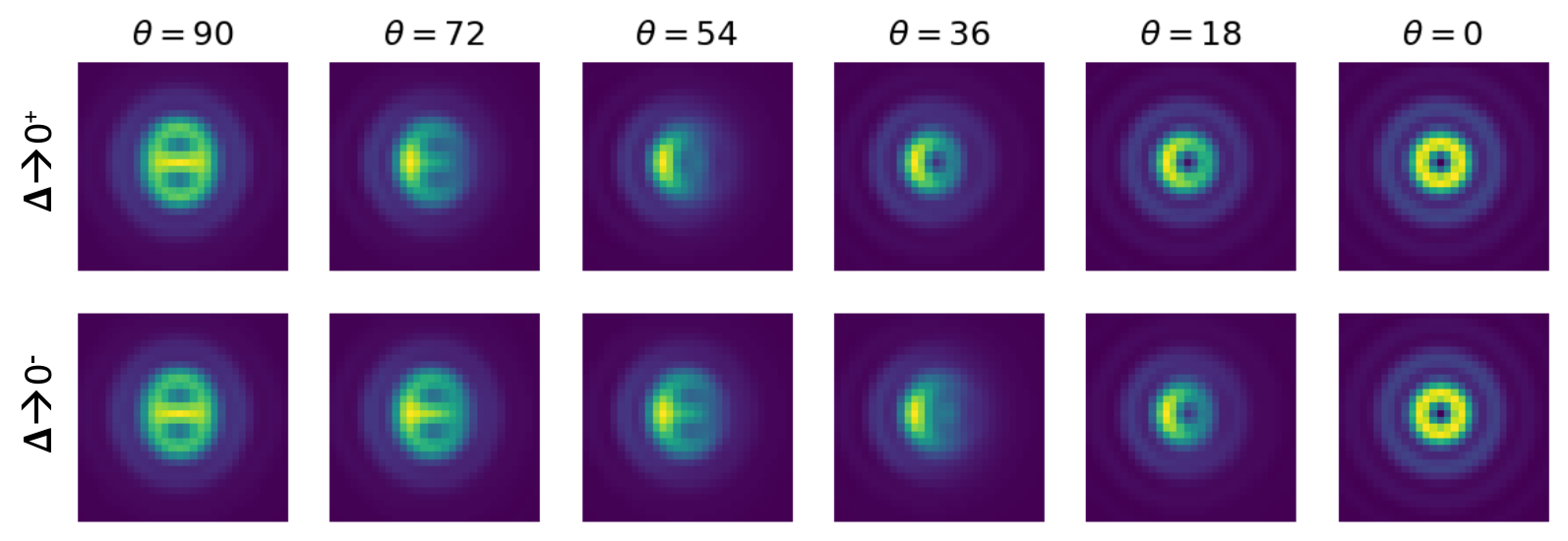}
\par\end{centering}
\caption{Sample radiation patterns for molecules simulated at various polar angles with 500~nm defocus at zero distance from the interface with the above-surface (top) and below-surface (bottom) prescription. The predicted shapes for a given dipole are inequivalent for these two prescriptions.  }\label{fig:Sample-Patterns}

\end{figure}

Examples of some molecular patterns predicted with these equations
are shown in Fig \ref{fig:Sample-Patterns}, top row. These are produced
for dipoles with orientation vectors
\begin{equation}
\mathbf{p}=(\sin\Theta\cos\Phi,\sin\Theta\sin\Phi,\cos\Theta),
\end{equation}
with $\Phi=0$ and varying $\Theta$. We see that the effect of varying
the polar angle $\Theta$ is a gradual shift from a circular
image near $\Theta=0$ (dipole vector perpendicular to interface)
to an asymmetrical one at $\Theta=90$ (dipole vector lying in the
plane of the interface). The effect of varying $\Phi$ is a straightforward
rotation of these patterns in the image plane. The precise shape of
the emission pattern depends on the defocus distance, refractive indices
and pixel size, but patterns like these can always be obtained using the
above calculations, which we henceforth refer to as the ``standard
method'', for any set of input parameters.

\section{The Surface Orientation Ambiguity}

There is a problematic ambiguity associated with the above result,
which forms the main subject of this paper, which we term the \textbf{Surface
Orientation Ambiguity}. A paradox is apparent when we consider same
limit as evaluated again, but now taken from the other side of the
interface, $\Delta\rightarrow0^{-}$ rather than $\Delta\rightarrow0^{+}$. This leads to an inequivalent
result for the $c$ coefficients, and hence for the image pattern,
\begin{equation}
c_{1}^{\Delta\rightarrow0^{+}}=\xi \,c_{1}^{\Delta\rightarrow0^{-}},\quad\quad c_{2}^{\Delta\rightarrow0^{+}}=c_{2}^{\Delta\rightarrow0^{-}},\quad\quad c_{3}^{\Delta\rightarrow0^{+}}=c_{3}^{\Delta\rightarrow0^{-}},\quad\quad \xi=\left(\frac{n_{1}}{n_{2}}\right)^2.
\end{equation}
Example images of the same molecules predicted under this second limiting
case are shown in Fig~\ref{fig:Sample-Patterns}, lower panels. A difference between the
above- vs below-surface zero distance solutions has been noted in
the past~\cite{Lukosz_III} in the context of the emission power of vertical dipoles. Its implications and resolution have not
been explored, to our knowledge. Resolving this issue is a matter
of both academic and also practical interest. 

\begin{figure}
\begin{centering}
\includegraphics[width=0.9\columnwidth]{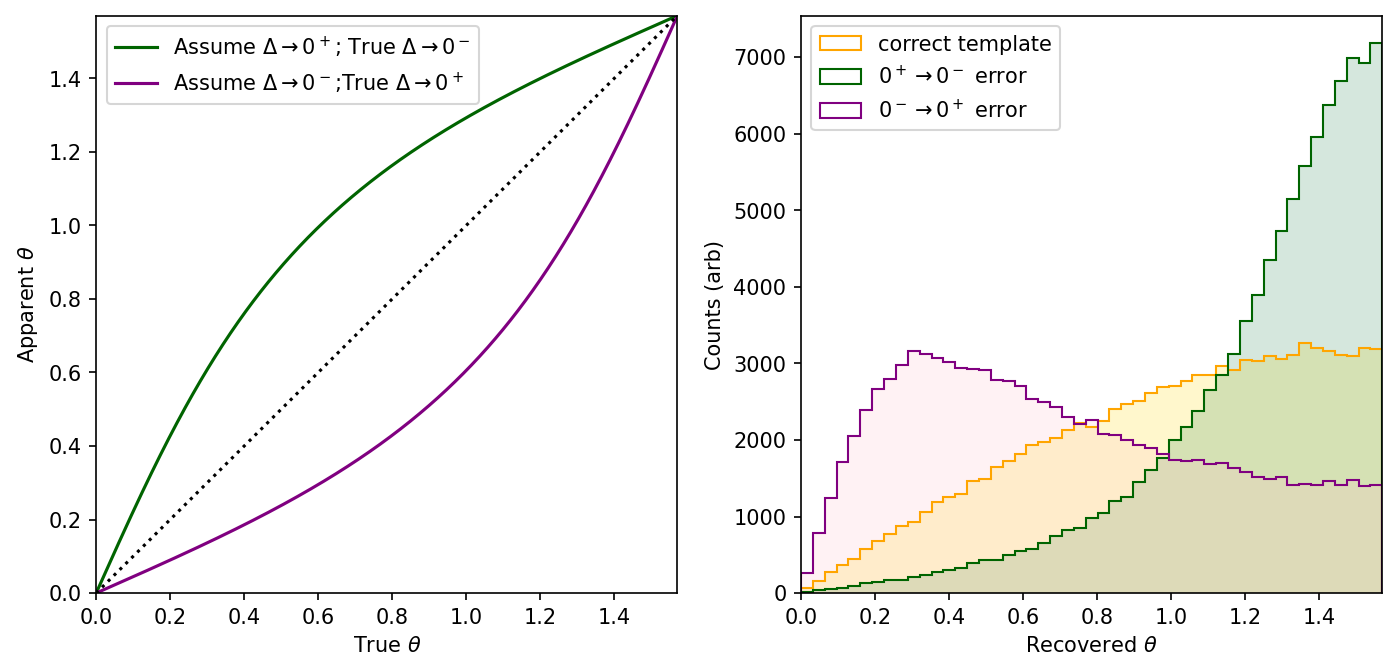}
\par\end{centering}
\caption{The effect of using one of the two limiting prescriptions if the other were correct is exactly equivalent to a systematic rotation of the dipole emission pattern. The left figure shows the true vs reconstructed angle for each case; the right figure shows how an isotropically distributed set of molecular orientations would be reconstructed in each case.~\label{fig:OrientationError}}
\end{figure}

Considering the Greens function of Eq. \ref{eq:GreensFunction-1},
we can see that the ambiguity discussed above has a clear observational
implication. If the correct answer were given by the $\Delta\rightarrow0^{-}$
solution but the $\Delta\rightarrow0^{+}$ one were assumed, the first
two columns of $\overleftrightarrow{\mathbf{G}}_{bfp}$ would be unaffected,
but the third column would be scaled by $(n_{1}/n_{2})^{2}$. This
means that a dipole with vector $\mathbf{p}=(p_{x},p_{y},p_{z})$
would be reconstructed as if it had unit vector $\mathbf{p}'=(p_{x},p_{y},\frac{n_{1}^{2}}{n_{2}^{2}}p_{z})$.
Since orientation experiments are typically concerned with image spatial
shape rather than brightness, an apparent rotation of the reconstructed
polar angle $\Theta$ to $\Theta'$ would be observed, related by
\begin{equation}
\Theta'=\mathrm{arctan}\left[\xi\tan\Theta\right].\label{eq:CorrectAngle}
\end{equation}
Importantly, the pattern produced would be a valid  prediction for the original set of templates,
set, but for a molecule with the incorrect polar angle. For situations
of practical interest, this is an effect of much relevance, since it can lead
to important interpretational errors. Fig.~\ref{fig:OrientationError} shows the correction
to the polar angle that would be expected at a glass-air interface
($n_{2}=1,n_{1}=1.5$) with either combination of one limit being
assumed and the other being true. Fig.~\ref{fig:OrientationError} right shows the
distribution of polar angles that would be reconstructed for an isotropic
distribution under these mis-reconstruction scenarios. That there is a striking similarity of the mis-reconstructed curve
to, for example, to Figure 4 of ref.~\cite{TenopalaCarmona2023Orientation}, which purports to observe
preferential orientation directions of molecules near surfaces as
opposed those in the bulk. 

A molecule sitting at an interface is in contact with both media, and not unambiguously situated in either material. As such, it is unclear which model (if either) should be taken as valid for this scenario. Furthermore, there is an unphysical
inconsistency implied in the transition region. A minute shift in
the position of the emitting dipole across the boundary, by say, $10^{-6}$
wavelengths of the emitted light, would be predicted to lead to a
dramatic and discrete change in the image pattern. Although both solutions
are exact and derived from Maxwells equations without any apparent
approximations, it appears that they must be missing an important
ingredient not only at the $\Delta\rightarrow0$ limit, but also at
short yet finite distances to the interface. 

It is interesting to consider both what is missing, and at what
distance the standard treatment should be expected to fail.
By construction,  it accounts for all near-field as well
as far-field optical effects, and the above- and below- patterns are each
essentially invariant for all distances $|\Delta|<\lambda/10$. As
such, it does not appear to be the wavelength of light that sets the
relevant distance scale for validity. We also cannot identify any
cocktail of atomic / quantum effects at the surface as the scapegoat,
nor blame microscopic roughness or other imperfections of the interface.
Such effects may surely further complicate the calculation, but the surface orientation ambiguity is present even in our idealized system with
a flat and smooth dielectric interface.  Indeed, a similar question could be posed for
a macroscopic, classical system, such as a transmitting radio-antenna floating at an air-water boundary.  For such a system one can consider raising and lowering it across the boundary
with the expectation of a continuum of emission patterns during this transition. This analogy
 makes it clear that the issue at hand is associated
with electrodynamics, rather than with microscopic surface details. In the following section we will proceed to resolve this ambiguity,
showing that the correct solution for the zero distance limit is neither
of the above limiting cases, but can be obtained and smoothly
connected to the analytic solutions.

\section{Resolution of the Surface Orientation Ambiguity Using Finite-Volume
Dipoles}

In practice, the radiating element at a surface is in contact with
both materials, but is neither within one of them or the other. Maxwells
equations for dielectric media are approximations, that work well
whenever a current element can be considered as existing at a location
with well defined dielectric and magnetic constants. We recall that
the dielectric constant is defined in order to conveniently encode
the contribution of the local polarization field \textbf{P }of a material
so that the total electric field is given by
\begin{equation}
\mathbf{D}\equiv\epsilon_{0}\left(\mathbf{E}+\mathbf{P}\right)\sim\epsilon_{0}\epsilon_{r}\mathbf{E}.
\end{equation}

At the surface itself, the polarization vector $\mathbf{P}$ is a
nontrivial function of both media and in fact does not actually point
in the same direction as \textbf{E}, due to the different polarizibilities
of the two half-spaces. As such, there is no well-defined scalar $\epsilon_r$
at the location of an interfacial emitting molecule. This prohibits the use of
Maxwells second equation for a point-like current element radiating
at or very near this boundary, since by construction this equation
requires values to be specified for $\mu_{r}$ and $\epsilon_{r}$
at the location $\mathbf{r}$ of the current, 
\begin{equation}
\frac{1}{\mu_{0}\mu_{r}(\mathbf{r})}\nabla\times\mathbf{B}(\mathbf{r})=\mathbf{j}(\mathbf{r})+\epsilon_{0}\epsilon_{r}(\mathbf{r})\frac{\partial\mathbf{E}(\mathbf{r})}{\partial t}.
\end{equation}.

This inconsistency is at the root
of the surface orientation ambiguity, and we can avoid this problem
if we instead consider a dipole that occupies a finite volume and
does not purport to occupy space within one material or the other.
For example, we can model the radiating dipole in terms of
as a surface current in a conductor that is in contact with both media.
Field matching can then be performed at all surfaces without ambiguity,
enabling prediction of the radiated electrical power.

\begin{figure}
\begin{centering}
\includegraphics[width=0.8\columnwidth]{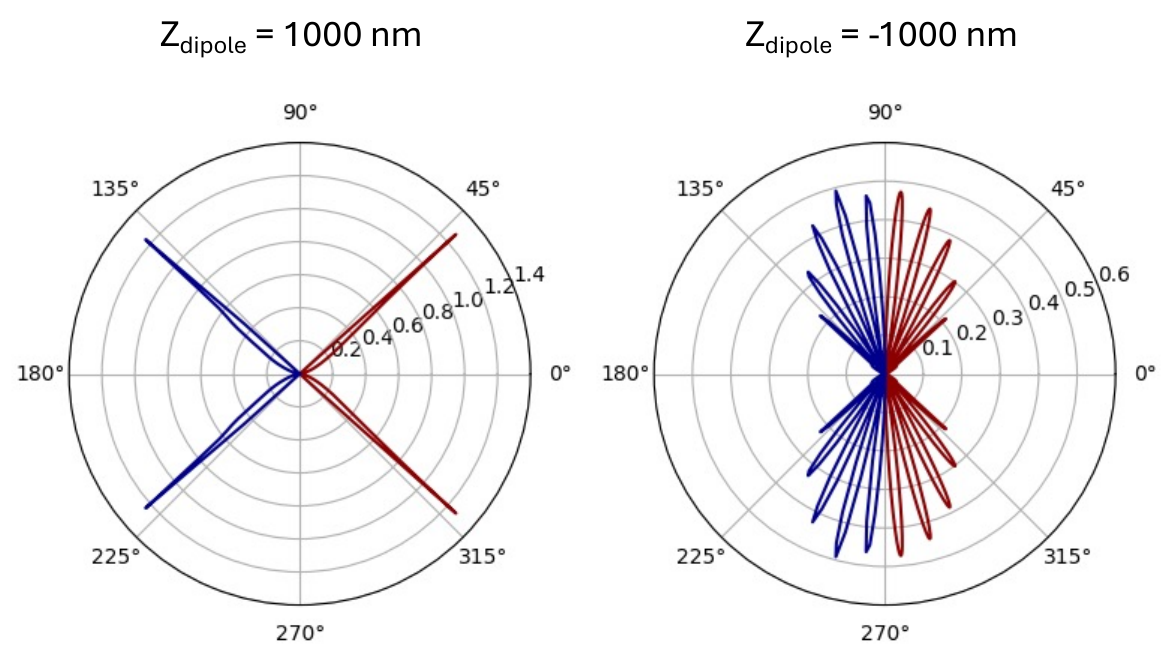}
\par\end{centering}
\caption{Radiation patterns for the far-from-surface (1000~nm) above- and below- cases in both the traditional theoretical calculation (red) and the finite-element method (blue), for a vertical dipole. There is good agreement between the two methods at these larger distances. }\label{fig:Caption}
\end{figure}

To this end, one can use finite element analysis software to
to predict the relevant radiation patterns. To simulate such patterns we have used the COMSOL multiphysics package. A small conducting sphere has surface currents driven at
the specific frequency corresponding to a vacuum wavelength of 550~nm. The sphere is placed within a geometry representing the interface
between two dielectric materials, and Maxwells equations are numerically solved
everywhere in space around the dipole, extending to a radius large enough
to predict the far-field radiation. 

To simplify the nature of the calculation that must be performed,
we note that the electric field can always be decomposed into a sum
of fields from the component of the dipole perpendicular to the interface
and another component parallel, 
\begin{equation}
\mathbf{p}=p_{z}\hat{\mathbf{z}}+\mathbf{p}_{\perp},
\end{equation}
and the electric fields generated by these components are additive,
by virtue of the linear nature of the Greens function,
\begin{equation}
\mathbf{E}_{bfp}=p_{z}\overleftrightarrow{\mathbf{G}}\hat{z}+\overleftrightarrow{\mathbf{G}}\mathbf{p}_{\perp}.
\end{equation}

The far electric field contribution associated with the $\hat{z}$ component
of the dipole depends only on $c_{1}$, whereas that associated
with the $\hat{x}$ and $\hat{y}$ components depends only on $c_{2}$
and $c_{3}$. There is no ambiguity in the standard calculation of
$c_{2}$ and $c_{3}$ across the boundary, and as such, we may restrict our attention
in these simulations to only the $\hat{z}$-oriented components. The
relevant electric fields can be derived from a $\hat{z}$-oriented
dipole, using axial symmetry to simplify the simulation geometry.
These components can then be added to the fields calculated using
the standard approach for the transverse dipole, yielding the result
needed for an arbitrary oriented emitter.

According to the standard calculation, the power radiated into angle
$\theta,\phi$ for a vertical dipole ($p_{z}=1)$ is given by~\cite{Novotny_Hecht_2012}
\begin{equation}
\frac{P(\theta,\phi)}{P_{d}}=\frac{3}{8\pi}\frac{n_{o}}{n_{d}}\sin^{2}\theta\,|c_{1}(\theta)|^{2}\label{eq:Power},
\end{equation}
where $n_{d}$ is the refractive index where the dipole sits, and
$n_{o}$ is the refractive index of the material lying in the $\theta,\phi$
direction. $c_{1}$ is the appropriate choice of $c$ function for
the half-space corresponding to angle $\theta$. Eq.\ref{eq:Power}
is provided as a ratio to $P_{d}$, the power of the same dipole in
a uniform medium of refractive index $n_{d}$. This denominator also
has a dependency on refractive index, as

\begin{figure}
\begin{centering}
\includegraphics[width=0.8\columnwidth]{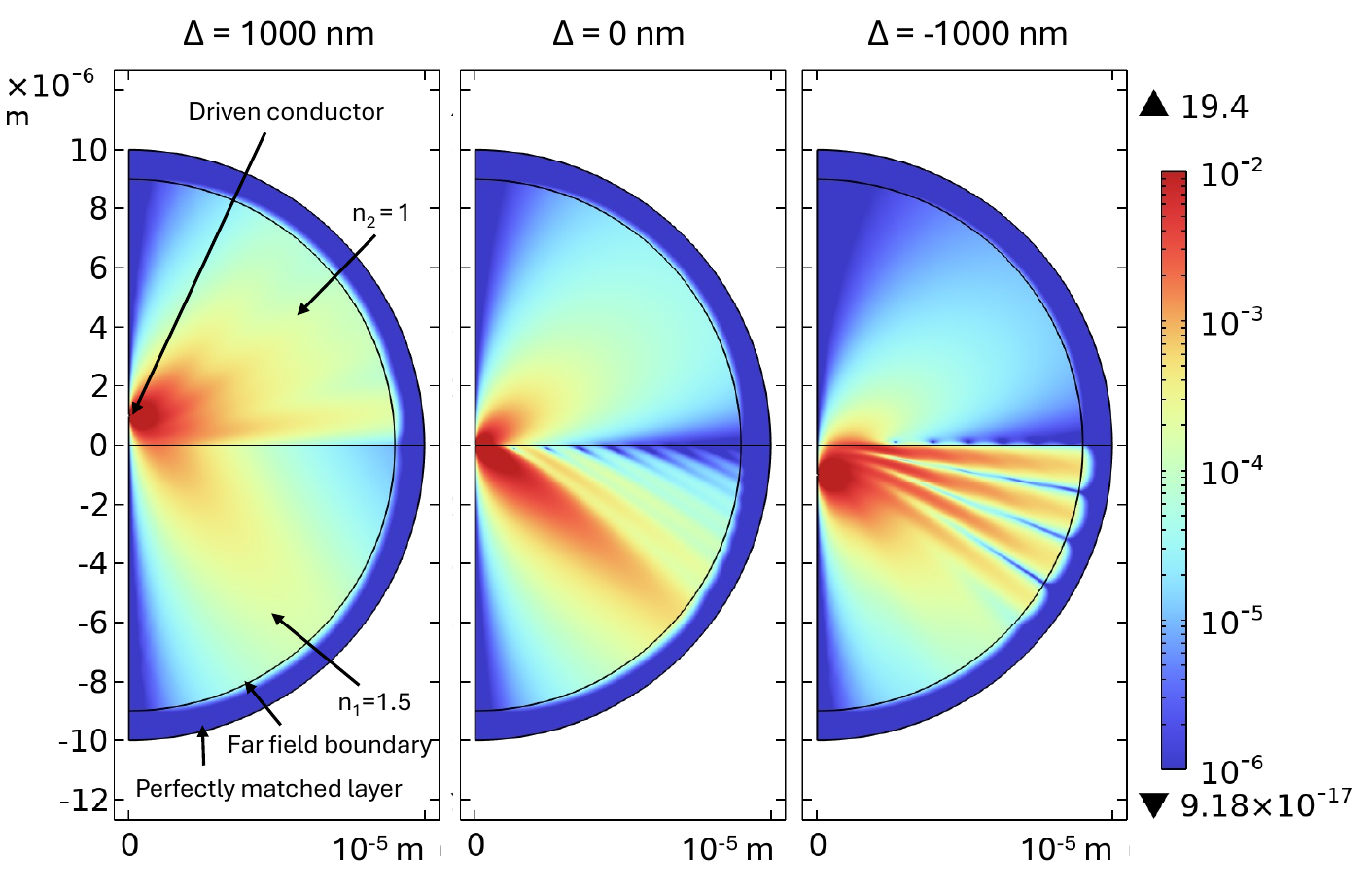}
\par\end{centering}
\caption{Maps of the radiated power in the near- and far-fields for finite-sized oscillating spherical current elements simulated above surface (left), at-surface (center) and below-surface (right). In these cases the current element does not exist inside either medium, instead occupying a finite volume and radiating energy into the dielectric system. The radiation pattern is measured at the far-field boundary. In these images, a smaller simulation volume (radius 10~$\mu$m) is used to better illustrate the near-field behavior. Subsequent simulations in this work use a larger simulation volume of 20~$\mu$m to ensure full convergence of the far field behavior.}\label{fig:2D-maps}
\end{figure}

\begin{equation}
P_{d}=\frac{n_{d}}{4\pi}\frac{\omega^{4}}{3c^{3}}.
\end{equation}
The total power, which will later serve as our proxy for the normalization
of $c_{1}$, is the integral of $P(\theta,\phi)$ over solid angle
in the lower half space,
\begin{equation}
\mathcal{{\cal P}}=\frac{1}{16\pi}\frac{\omega^{4}}{c^{3}}n_{o}\int d\theta\sin^{3}\theta\,|c_{1}(\theta)|^{2}.
\end{equation}

We first verify that this system reproduces the expected radiation
pattern for a point dipole at relatively large distances between the
dipole and the surface. Figure \ref{fig:Caption} shows the radiation
patterns from the analytic calculations described above and the numerical
simulations with a finite dipole at distances of 1000 nm above and
below the surface. The far-field radiation pattern is obtained by
solving Maxwells equations over the full geometry and finding the
power crossing a far-field surface near the geometry edge, and applying
perfectly matched layer around the boundary of the simulation geometry
to avoid radiation scattering backward from the outer boundaries of
the simulation volume. The complete solutions showing both the near-field
and far-field power components of the magnitude of the Poynting vector
are shown for above- below- and at-surface simulations in Fig. \ref{fig:2D-maps}.
For ease of viewing, the illustrative plots of Fig. \ref{fig:2D-maps}
show results obtained using a simulation domain of $10\,\mu$m radius,
though notably in quantitative studies a somewhat larger geometry
of $25\:\mu$m is used, to ensure the radiation pattern in the far
field is well converged.   The shape of the radiation pattern for the on-surface simulation was
also compared to the standard solution in both $\Delta\rightarrow0^{+}$
and $\Delta\rightarrow0^{-}$ limits and found in good agreement. The key question for present purposes concerns the normalization of the $c_{1}$ function.

\begin{figure}
\begin{centering}
\includegraphics[width=1.0\columnwidth]{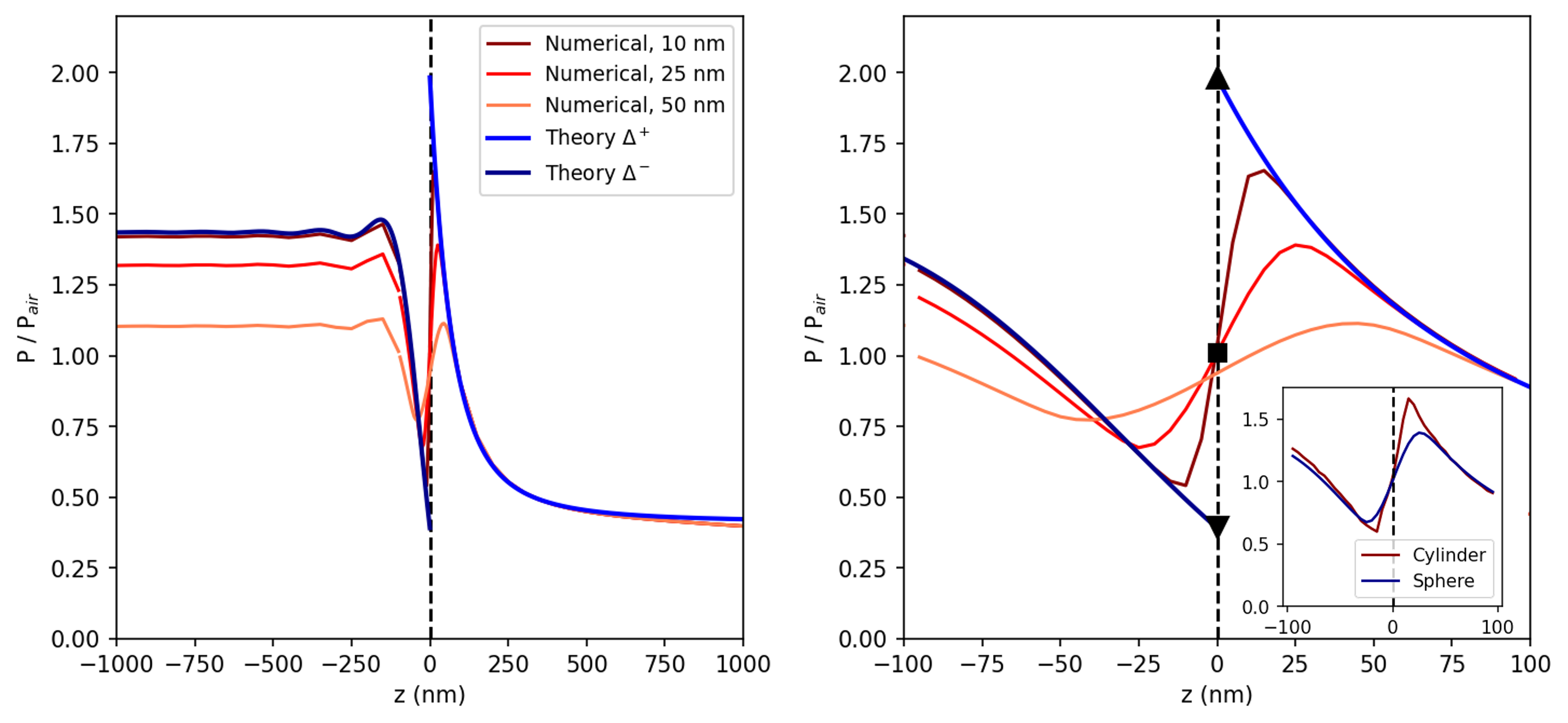}
\par\end{centering}
\caption{Radiated power from a vertical dipole for both the traditional calculation and for finite element simulations, over a wide (left) and narrow (right) range of distances.  The zero-distance predictions are shown as black markers on the right figure.  For finite-sized dipoles the simulations converge to a consistent prediction of the radiated power at zero-distance from surface, as long as the emitter dimensions are much smaller than the wavelength of the light (square point). The inset figure shows simulations for spherical and cylindrical emitters.  Some dependence on the current element shape is observed when the distance to the surface is comparable to the dipolar dimensions, though converging to the same zero-distance solution is observed.}\label{fig:Main-results-figure.}
\end{figure}

Figure \ref{fig:Main-results-figure.} shows the results of simulations
of small but finite vertical dipoles placed at different distances
from a dielectric interface, in terms of total power radiated into
the downward half-space. Each curve is normalized against a calculation
for the same shape of dipole simulated in a uniform medium of air.
Also shown is the theoretical curve from the calculations described
above. The power emitted by a vertical dipole $\mathcal{P}_{\mathbf{p}\parallel\mathbf{z}}^{\Delta\rightarrow0^{+}}$
is proportional to $c_{1}^{2}$, and as such, the power ratio for
the asymptotic analytic solutions is
\begin{equation}
\frac{\mathcal{P}_{\mathbf{p}\parallel\mathbf{z}}^{\Delta\rightarrow0^{+}}}{\mathcal{P}_{\mathbf{p}\parallel\mathbf{z}}^{\Delta\rightarrow0^{-}}}=\left(\frac{c_{1}^{\Delta\rightarrow0^{+}}}{c_{1}^{\Delta\rightarrow0^{-}}}\right)^{2}=\left(\frac{n_{1}}{n_{2}}\right)^{4}.
\end{equation}

For $n_{1}=1$ and $n_{2}=1.5$ this ratio is approximately 5.06,
and it can be observed on Fig \ref{fig:Main-results-figure.}, right
as the ratio of intersections of the two analytic models with the
x=0 axis, marked with upward and downward point triangles. The finite size dipole models, on the other hand, produce
smooth curves without discontinuities. We observe the numerical solutions
tending toward the analytic ones as the dipole radius tends toward
zero. Nevertheless, for all cases where the dipole is significantly
smaller than the wavelength of the light (in this example, 550 nm),
there is a unique and convergent prediction for the on-surface
power, $\mathcal{P}_{\mathbf{p}\parallel\mathbf{z}}$. This value
is intermediate between the two theoretical predictions discussed
earlier, marked on Fig~\ref{fig:Main-results-figure.}, right, corresponding to approximately $\frac{\mathcal{P}_{\mathbf{p}\parallel\mathbf{z}}}{\mathcal{P}_{\mathbf{p}\parallel\mathbf{z}}^{\Delta\rightarrow0^{-}}}\sim2.6$ for this combination of material parameters.  The shape of the emitting dipole has some impact for the very-near-surface solutions, but almost no effect at the $z=0$ limit. For example Fig.~\ref{fig:Main-results-figure.} right shows an inset panel comparing the near-surface results for a cylindrical and spherical dipole current density.  Their zero-distance solutions are equivalent at the 2\% level, which is consistent with the expected precision of these simulations.

For larger dipole radii, and particularly as the dipole dimensions
begin to become comparable with the wavelength of light, more complex
geometrical effects are observed. For dipoles larger than around 50~nm, even the zero-distance prediction begins to deviate from the short-dipole
limit, assuming a smaller value for the radiated power then predicted
for a small dipole. There is also a significantly different prediction
for light radiated into the upper vs lower half spaces compared with
the point-dipole approximation, which is a consequence of the interference
of electric field contributions emanating from different locations
across the emitter. While these observations are of academic interest,
they are irrelevant to the case of single molecules radiating near
dielectric surfaces, since in these scenarios the physical size of
the emitter is always far shorter than the wavelength of the fluorescence
light and such form factor effects can be neglected.

\begin{figure}
\begin{centering}
\includegraphics[width=1.0\columnwidth]{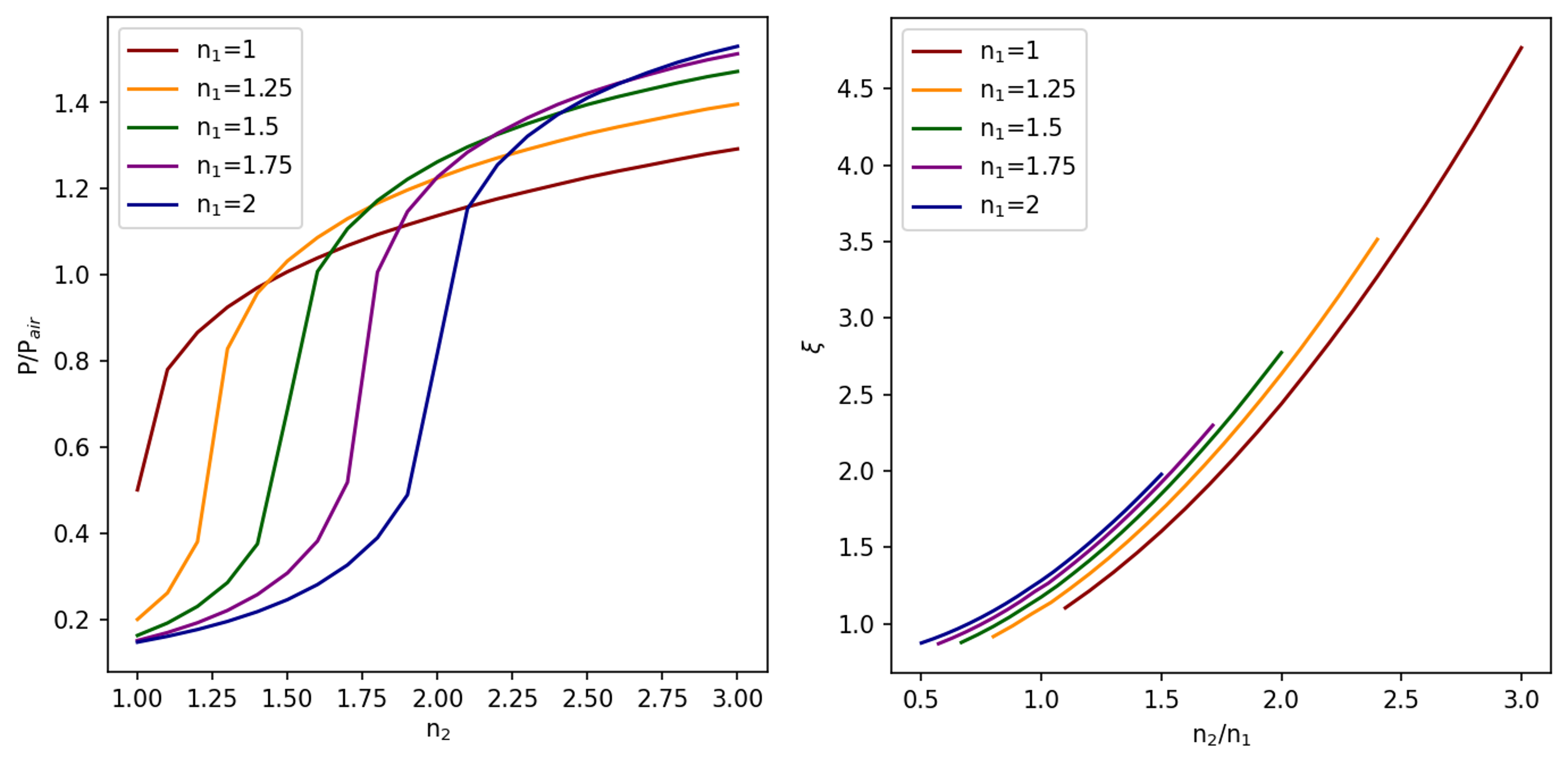}
\par\end{centering}
\caption{The power emitted by an on-surface vertical dipole depends on the refractive indices of both media (left).  This information is used to extract the required factor $\xi$, which can be used to either correct reconstructed angles for measurements made using the traditional approach, or equivalently to generate properly oriented emission templates for fitting data.}\label{fig:RefIndices}
\end{figure}

For the case with refractive indices n$_1$=1 and n$_2$=1.5, the correction parameter for an on-surface dipole relative to the $\Delta\rightarrow 0 ^+$ protocol has been found to be $\xi=\sqrt{2.6}\sim1.6$.  For the more general case, it is necessary to find the appropriate value of $\xi$ for other combinations of refractive indices. Fig.~\ref{fig:RefIndices} reports
calculated power ratios (left) and $\xi$ values (right) for several other combinations, spanning the range of interest to optical fluorescence microscopy. It is observed that there is a near-universal scaling with n$_2$/n$_1$, and that the correction is larger for more strongly mismatched indices.  A common power ratio of 0.5 for $n_1=n_2$ reflects that for a uniform medium, half the power is emitted into the downward half-plane toward the imaging system, as expected.

With the power emitted by the finite but small vertical dipole on a surface now understood, there is a clear prescription for finding the orientation of any dipole based on its defocused image. The horizontal components are unambiguously predicted by the standard protocol, and an arbitrary dipole field can be
constructed by superposing the electric field from a vertical dipole
with that from a horizontal one. As such, the correct molecular emission patterns
can then be found using the prescription $c_{1}=\xi c_{1}^{\Delta\rightarrow0^{+}}$ when generating fitting templates, or equivalently, by reconstructing using the standard method and then using Eq.~\ref{eq:CorrectAngle} to correct the polar angle, using an appropriately calculated value for $\xi$.

\section{Conclusions}

We have re-analyzed the problem of an electric dipole emitter radiating at the interface of two dielectric media, with particular focus on the case where the dipole is situated directly at the interface. The traditional calculations of the expected radiation pattern lead to an ambiguity, because the limiting form of the angular spectra at zero distance from the interface does not converge to a common prediction for above- vs below-interface dipoles.  This ambiguity can be resolved using finite element modeling with finite-sized current elements as a radiation source, leading to an unambiguous prediction for at-surface dipoles when the size of the emitter is smaller than the wavelength of emitted light.

These results have important implications for single molecule orientation measurements.  Use of the traditional ``above-surface'' calculation for the radiation pattern will lead to molecular orientation measurements that are systematically rotated into the plane of the interface.  This may have biased past works that have attempted to quantify molecular orientations for on-surface conditions, including for spin-coated or vapor-deposited single molecules. There are also possible implications for experiments that study emitters in self-assembled monolayers.  

We have provided a prescription for correcting the measured orientations for on-surface molecules that are in direct contact with both media based on knowledge of the two refractive indices.  The required correction becomes larger for large refractive-index ratios, scaling approximately as $n_2/n_1$ for modest values.

Although the zero-distance solution is shown to be unambiguous, caution should be still taken in applying this prescription to data.  While the radiation pattern converges to fixed shape and intensity as the emitter becomes much smaller than the fluorescence wavelength, the transition from the positive-like to negative-like solution becomes increasingly sharp.  This means that a very small separation from the surface may still change the angular response significantly. At this scale there are also additional effects, including surface roughness and non-parallel induced polarization vector of the interfacial fields that can become relevant for very-near-surface molecules. Owing to these complications, a direct experimental verification of the orientation patterns of at-surface emitters appears strongly motivated. Such an experiment would require preparation of molecular dipoles in well-defined orientations at the surface, a challenging but perhaps not insurmountable problem.  Such an experiment could shed further light on the effects discussed in this work, perhaps lifting the remaining ambiguities associated with at-surface orientation measurements.

\section*{Acknowledgements}

This research was supported by the US Department of Energy DE-SC0019054  and DE-SC0019223 (BJPJ and ED); the US National Science Foundation under award number NSF CHE 2004111 (FWF), the European Research Council (ERC) under Grant Agreement No.\ 951281-BOLD (JJGC and ME) and the Department of Science, Universities, and Innovation of the Basque Government PRE$\_$2024$\_$2$\_$0260 (ME).  The authors thank the NEXT collaboration for productive discussions during development of this work.

\section*{Disclosures}
The authors declare no conflicts of interest.
\bibliography{Dipole}
\end{document}